\documentclass[aps, pre, onecolumn, amsmath, superscriptaddress,showkeys,showpacs]{revtex4-1}
\usepackage{commath}
\usepackage[innercaption]{sidecap}
\usepackage{amssymb}
\usepackage{graphicx}
\usepackage{graphicx,epstopdf}
\usepackage{xcolor, soul}
\usepackage{dcolumn}
\usepackage{bm}
\usepackage{mathrsfs} 
\usepackage{amsmath} 
\usepackage[colorlinks=true,linkcolor=blue,citecolor=blue]{hyperref}%
\usepackage{fontenc}
\usepackage{float}
\usepackage{amsthm}
\usepackage{subfigure}
\usepackage{color}
\usepackage{ragged2e}
\usepackage{enumerate}
\begin{document}
\title[Running Title]{Emergence of mixed mode oscillations in random networks of diverse excitable neurons: the role of neighbors and electrical coupling} 
\author{Subrata Ghosh$^{1}$, Argha Mondal$^{1}$, Peng Ji$^{5,*}$, Arindam Mishra$^{2}$, Syamal K. Dana$^{2,3}$, Chris G. Antonopoulos$^{4}$, Chittaranjan Hens$^{1,*}$}
\affiliation{$^{1}$Physics and Applied Mathematics Unit, Indian Statistical Institute, Kolkata, India\\
$^{2}$Centre for Mathematical Biology and Ecology, Department of Mathematics, Jadavpur University, Kolkata, India\\
$^{3}$Division of Dynamics, Faculty of Mechanical Engineering, Lodz University of Technology, Lodz, Poland\\
$^{4}$Department of Mathematical Sciences, University of Essex, Wivenhoe Park, UK\\
$^{5}$The Institute of Science and Technology for Brain-inspired Intelligence, Fudan University, Shanghai, China}
\def\corrAuthor{Chittaranjan Hens and Peng Ji}
\def\corrEmail{chittaranjanhens@gmail.com, pengji@fudan.edu.cn}
%


\begin{abstract}
\noindent In this paper, we focus on the emergence of diverse neuronal oscillations arising in a mixed population of neurons with different excitability properties. These properties produce mixed mode oscillations (MMOs) characterized by the combination of large amplitudes and alternate subthreshold or small amplitude oscillations. Considering the biophysically plausible, Izhikevich neuron model, we demonstrate that various MMOs,  including MMBOs (mixed mode bursting oscillations) and synchronized tonic spiking appear in a randomly connected network of neurons,  where a fraction of them is in a quiescent (silent) state and the rest in self-oscillatory (firing) states. We show that MMOs and other patterns of neural activity depend on the number of oscillatory neighbors of quiescent nodes and on electrical coupling strengths. 
Our results are verified by constructing a reduced-order network model and supported by systematic bifurcation diagrams as well as for a small-world network. 
Our results suggest that, for weak couplings, MMOs appear due to the de-synchronization of a large number of quiescent neurons in the networks. The quiescent neurons together with the firing neurons produce high frequency oscillations and bursting activity. The overarching goal is to uncover a favorable network architecture and suitable parameter spaces where Izhikevich model neurons generate diverse responses ranging from MMOs to tonic spiking.
\end{abstract}
\maketitle
\section{Introduction}

Diverse spiking oscillations and bursting phenomena of electrical activity in single neuron or neuronal networks play an important role in information processing and transmission across different brain areas \citep{connors1990intrinsic,izhikevich2003simple,izhikevich2004model,coombes2005bursting,izhikevich2007dynamical,antonopoulos2015brain,Ma2017,teka2018,mondal2018diverse,antonopoulos2019evaluating}. 
The underlying mechanism of signal processing in neurons depends on the variations of membrane voltages called spikes \citep{izhikevich2003simple,izhikevich2004model,izhikevich2007dynamical}. The complexity of spikes or trains of spikes can be controlled by external stimuli, e.g. by injected electrical currents. In a common scenario, a bunch of spikes (called a burst) may emerge in the activity of single neurons or in neural populations \citep{izhikevich2000neural,coombes2005bursting,Constantinou2016,Zeldenrust2018}. Such oscillatory patterns of membrane voltages can be modeled mathematically by biophysical dynamics (with realistic parameters) such as the (un)coupled Izikevich neuron model \citep{Khoshkhou2018}, described in the next section.
Our goal is to study the firing and collective activities of coupled neurons in an environment of heterogeneous excitabilities. Neural networks support functional mechanisms within brain areas. For example, such diverse groups of neurons in the cortex are responsible for many complex neuronal mechanisms \citep{izhikevich2000neural,izhikevich2004model,izhikevich2007dynamical}. 
 
Most of the neurons are excitable, i.e., they show quiescent behaviour however, they can also fire spikes when they are stimulated by input stimuli. In neural computations, the neurons continue to fire a train of spikes when there is an input by injecting a pulse of dc current and this is called tonic spiking. There exist different types of spiking patterns depending on the nature of the intrinsic dynamics. Bursting follows a dynamic state in a neuron where it repeatedly fires discrete groups or bursts of spikes, i.e., when the activity alternates between a quiescent state and repetitive spiking (a bunch of spikes appear together). This might be regular or chaotic, depending on the dynamics of the system and excitabilities or couplings \citep{izhikevich2000neural,izhikevich2004model,izhikevich2007dynamical}. Apart from spiking and bursting activities, one of the interesting complex firing patterns {\color{red}} emerge from the activity of neurons is the mixed-mode oscillations (MMOs) \citep{brons2008introduction,desroches2012mixed,bacak2016mixed}, what is the main focus here. In MMOs, the oscillations are distributed with different amplitudes where the firings alternate between large and small amplitude oscillations \citep{brons2008introduction} (i.e., the so called $LAO$s and $SAO$s, respectively)  reflecting different rhythmic activities such as locomotion or breathing \citep{bacak2016mixed}. The multiple time scales (e.g. fast potassium channels with slow kinetics \citep{ghaffari2015effect}) of voltage variables or controlled noise can induce MMOs in neuronal systems \citep{muratov2008noise,upadhyay2017mixed}. MMOs were first observed in chemical reaction systems \citep{ostwald1900periodische}. They were also observed in Belouzov-Zhabotinsky reactions \citep{schmitz1977experimental,showalter1978modified,brons1991canard}, calcium dynamics and electrocardiac systems \citep{kummer2000switching,rotstein2006localized}.
We note that, from a dynamical perspective, the generation of MMOs can be analyzed through the canard phenomenon \citep{drover2004analysis,eckhaus1983relaxation,rubin2008selection} and also via homoclinic bifurcations \citep{Dana2010}. 
Krupa et al. \citep{krupa2008mixed} analyzed the mechanism of MMOs in a two-compartmental model of dopaminergic neurons in the mammalian brain stem. 
To investigate the generation of MMOs in a self-coupled, FitzHugh-Nagumo model, Desroches et al. \citep{desroches2008mixed} developed a computational method 
and Guckenheimer \citep{guckenheimer2008return} examined how chaotic dynamics and MMOs arise near folded nodes and folded saddle-nodes on slow manifolds.
Vo et al. \citep{vo2010mixed} demonstrated that MMOs can generate a type of bursting that can be reflected in a biophysical model of pituitary lactotroph \citep{toporikova2008type}. 
MMOs were also observed in stellate cells of the medial entorhinal cortex (layer II) and Rotstein et al. \citep{rotstein2008canard} analyzed the mechanism of such patterns in a biophysical, conductance-based, model. Apart from MMOs, mixed-mode bursting oscillations (MMBOs) \citep{desroches2013mixed} were also observed when a bunch of spikes in a single burst appears with $SAO$s. In MMBOs, burst activity appears instead of single spikes within LAOs. Our study on network dynamics sheds more light on such interesting patterns.

In this paper, we explore the emergence of spiking and MMOs in a random network of diffusively coupled (through the membrane voltage variable) Izhikevich neurons in a backdrop of diverse excitabilities. The role of network structure and arrangement of mixed neural populations in the network are the main objectives for the study of the emergence of MMOs. 
In network neuroscience, researchers investigate the firing activities and collective patterns of neural activity where neurons are connected in a complex-network topology \citep{brons2008introduction,erchova2008rhythms,desroches2008mixed,Postnov2008multi,krupa2014weakly,Malagarriga2015,antonopoulos2016dynamic,borges2017spike,Khoshkhou2018,borges2020self}. For instance, a correlated synchronous firing appears in neuronal cells with the adaptive exponential integrate-and-fire model with excitatory-inhibitory synapses that can be associated with epileptic seizures \citep{protachevicz2019bistable}. Bittner \citep{bittner2017population} showed that balanced excitatory and inhibitory input currents in clustered (non-clustered) networks of neurons may reflect spiking activities in which inhibitory neurons share more coherent activities. Recently, MMOs have also been observed in pre-B\"otzinger complex networks \citep{bacak2016mixed} (a medullary region that controls breathing in mammals) in the presence of heterogeneous excitable parameters. In both studies, a three-coupled reduced model was proposed to understand the behavior of collective spiking patterns and the conditions for the emergence of $LAO$s and $SAO$s were studied.

 However, the role of network architecture and different excitabilities in the emergence of MMOs are not well understood. In this paper, we have affirmative answer to the question related to the emergence of MMOs. We reveal how such MMOs can be distinguished from other firing patterns, supported by their relevant biophysical significance \citep{golomb2014mechanism}. Moreover, the neurons in the paper are placed on the nodes of a random network and  transfer signals through its links. In the absence of coupling, the activity of the considered neuronal population reveals two types of dynamical states (or excitabilities), ranging from spike-bursting to subthreshold to quiescent states. The key question that arises here is the following: considering a mixed/heterogeneous neural population (neighboring neurons of self-sustained spiking neurons might have subthreshold oscillations), can we design a random network of neurons (with Poissonian neighbor node-degree-distribution) that will give rise to collective firings where subthreshold or quiescent neurons are compelled to show high amplitude activities? We want to uncover the coupling parameter space and the ratio of mixed populations where MMOs and fast tonic spiking behavior emerge. In this context, by mixed/heterogeneous neural population we mean that neurons with different excitability properties i.e., the non-identical neurons with different  firing patterns are connected in a complex network. At weak couplings and a diluted random network setting, we show that desynchronized subthreshold neurons exhibit MMOs. With the increase of the coupling, all subthreshold neurons fire in a mixed-mode state. In both cases, MMOs are not prominent in oscillatory neurons and eventually disappear as the coupling strength increases. Consequently, neural subpopulations emerge as synchronous clusters exhibiting tonic spiking behavior. For diluted random and homogeneous networks, where the electrical coupling strength is constant, we show that neighbors exhibiting self-sustained oscillations, determine the structural patterns of MMOs. Based on the synchronised cluster over a certain coupling range, we can reduce the random network to a low dimensional, reduced-order network, i.e., to two coupled oscillators which reflect and predict the diverse dynamical patterns that appear in the random network. Additional to the random network,  we have  validated our results in  small-world network of $500$ nodes. In particular, our results for both types of networks confirm that the emerging features observed in the random network can also be found in the small-world network.

The paper is organized as follows: in Sec. \ref{Biophysical model and network}, we describe the Izhikevich neuron model and discuss its dynamical properties. The model displays various electrical activities (i.e., different spiking and bursting patterns) for fixed parameter values and for a range of injected currents, $I$. Then, we investigate the dynamical behavior on a random network (see Subsec. \ref{network}) based on single Izhikevich neurons with various firing responses. In particular, we identify the parameter region and coupling strategy where MMOs and MMBOs exist, and analyze the transition phases of firing responses (Subsec. \ref{imp_spiking_neighbors} and \ref{imp_electricalcoup}). In Sec. \ref{reduce network}, the reduced-order network model is constructed to verify the results obtained for the random network. A bifurcation analysis is also performed to show the mixed mode states and other phases of oscillations. In Sec. \ref{MMO in SMW}, the MMOs are further tested in a small-world network.
Finally, we conclude our work in Sec. \ref{Conclusions}, followed by a discussion.

\section{Biophysical model and random network}
\label{Biophysical model and network} 
\subsection{Model description}
\label{Model} 
Our work focuses on the analysis of the complex dynamical behavior in the 2-dimensional nonlinear Izhikevich model that captures neuronal membrane voltages \citep{izhikevich2003simple,izhikevich2004model}. It produces spiking and bursting patterns distributed over a range of parameter values. It is a biophysically plausible and computationally efficient mathematical model that takes into account continuous spike generation and a discontinuous resetting process following the spikes. It has two state variables; the membrane voltage, $v$ and recovery variable, $u$, which measure the activation of K$^+$ and inactivation of Na$^+$ ionic currents, respectively. The dynamical activity of an Izhikevich neuron is captured by the set of equations
\begin{align}
\dot{v}&=0.04v^2 + 5v +140 -u +I, 
\label{Varv} \\
\dot{u}&=a(bv-u),
\label{Varu} 
\end{align}
with an after-spike resetting constraint, i.e., when the membrane voltage $v$ reaches a peak value $v_{pk}$, the following relation is applied: if $v\geq v_{pk} (=30)$, 
then $v\gets c$ and $u\gets u+d$. The parameters $a$, $b$, $c$ and $d$ are dimensionless. The resting potential ranges in the interval $-70$mV to $-60$mV and depends on $b$ that indicates the sensitivity of $u$ to the subthreshold fluctuations of the membrane potential, $v$. The parameter $a$ measures the timescale of the recovery variable $u$. The parameters $c$ and $d$ control the after-spike reset value of $v$ and $u$, respectively, caused by fast high-threshold K$^+$ channel conductances and slow Na$^+$ and K$^+$ conductances. The function $(0.04 v^2 + 5v + 140)$ was derived using the spike initiation dynamics of a cortical neuron. The different suitable choices of parameters generate various types of oscillations, often found in neocortical and thalamic neurons \citep{connors1990intrinsic,gray1996chattering,izhikevich2000neural}. The initial conditions are set to $v=-63$ and $u=bv$. Synaptic currents or injected DC-currents are delivered via $I$. We consider a fixed parameter regime that produces different firings for a single Izhikevich neuron \citep{izhikevich2003simple,izhikevich2004model}, i.e., $a=0.1$, $b=0.2$ with reset parameters $c=-65$ and $d=8$, what we call set I.
We note that for $I<4$, the system of Eqs. \eqref{Varv} and \eqref{Varu} does not show any spiking or bursting behavior. Thus, the firing patterns can be obtained for $I\geq 4$. Simulations of the systems of ordinary differential equations were performed using the fourth-order Runge-Kutta method with a fixed time step of 0.01, as the simulation results with a smaller time step did not show any significant differences. Bifurcation diagrams of the deterministic dynamical model in the reduced-order network were computed using the MatCont software package \citep{Dhooge2003}.

\vspace{1.5cm}
\subsection{Formulation of the network of model neurons}\label{network}

We construct an Erd\H{o}s-R{\'e}nyi (ER) random network of $N=500$ nodes with average node-degree $5$. Then, we set up a mixed population of Izhikevich neurons to model neural activity on the nodes of the random network, where $70\%$ of them exhibit oscillatory behavior (self-sustained spiking oscillations, for $I=10$) as shown in Fig. \ref{TS_Spatio1}(b) (in blue) and $30\%$ are in quiescent states (for $I=3$), shown in Fig. \ref{TS_Spatio1}(b) (in red) by setting all the parameters in the tonic spiking condition (see set I). 
The system is coupled via the membrane voltage $v$ with a mean-field diffusive coupling. In particular, the equations of the $N$ coupled neurons ($i=1,2,\ldots,N$) in the network are described by
\begin{align*}
{\dot{v}_{i}}&=0.04v_{i}^2 + 5v_{i} + 140 - u_{i} + I_i + \frac{K}{\sum_{j=1}^{N} A_{ij}}\sum_{j=1}^{N} A_{ij} (v_{j} - v_{i}),\\
{\dot{u}_{i}}&=a(b v_{i} - u_{i}),
\end{align*}
with the constraint that if $v_{i}\geq30$, 
then, $v_{i}\gets c$ and $u_{i}\gets u_{i}+d$. $A$ is the adjacency matrix of the random network, $K$ the coupling strength and $S_i=\sum_{j=1}^{N} A_{ij}$ the degree of the $i$th node. We consider $I_1=\ldots=I_p=3$ where $\frac{p}{N}=0.3)$ and $I_{p+1}=\ldots=I_N=10$ where $q=1-\frac{p}{N}=0.7)$ that lead to the time evolution shown in Fig. \ref{TS_Spatio1}(a),(b). In the absence of coupling, the oscillatory nodes ($70 \%$) show desynchronized spiking and the rest of them ($30\%$) converge to fixed points (see spatiotemporal plot in Fig. \ref{TS_Spatio1}(c), where the inset is a zoom-in). With the increase of the coupling strength $K$,  the quiescent neural subpopulation exhibits different transitions to oscillatory behavior. Generally, for weak coupling, this subpopulation generates MMOs and subthreshold oscillations. One type of MMOs shows that between two consecutive $LAO$s, there exist two $SAO$s. Interestingly, other aperiodic MMOs may coexist in this subpopulation. Interspike intervals ($ISI$) are not identical and the number of small amplitude spikes in $SAO$s within two large amplitude  spikes may vary in the entire signal. We have found three types of MMOs shown in Fig. \ref{TS_Spatio1}(e), randomly picked from the quiescent subpopulation in which the average interspike intervals, $\langle ISI\rangle$, differ significantly. We will analyze such mixed MMOs behavior and variation of $SAO$s between $LAO$s in the next subsections. This study unveils the generation and annihilation of MMOs within a subpopulation of neurons. We note that, the oscillatory subpopulation shows almost coherent tonic spiking (Fig. \ref{TS_Spatio1}(d)). The spatiotemporal plot of all nodes is shown in Fig. \ref{TS_Spatio1}(f), where quiescent nodes are desynchronized (a zoom-in is shown on the right). With further increase of the coupling ($K=0.4$), the quiescent subpopulation exhibits MMOs, however the number of $LAO$s between two spikes is considerably decreased. The distance between two consecutive spikes is also decreased compared to the previous coupling case, therefore, $\langle ISI\rangle$ is also decreased (see Fig. \ref{TS_Spatio1}(h), where two randomly chosen nodes have been depicted in the panels of the figures. Interestingly, the oscillatory subpopulation remains in the same firing regime and the network shows asynchronous behavior (Fig. \ref{TS_Spatio1}(g),(i)) for all nodes. Finally, for $K=1$, the complete population switches to tonic spiking (Fig. \ref{TS_Spatio1}(j),(k),(l)) with almost identical $\langle ISI\rangle$, and the two subpopulations form two clusters when they are separately synchronized.

\subsubsection{MMOs in the quiescent subpopulation: impact of spiking neighbors of quiescent nodes}\label{imp_spiking_neighbors}

Here, we elaborate on the quiescent population and on several coexisting MMOs that emerge. Figure \ref{Network_node}(a) shows the network structure with a mixed population (spiking neurons are shown with blue filled circles and quiescent nodes with red filled circles). We first observe the emergence of MMOs in the quiescent nodes at weak coupling. At $K=0.3$, we have isolated three red nodes with different neighbor distributions. The red node (left) with $7$ neighbors shows MMOs in which three large amplitude  spikes exist within $100$ time units (see Fig. \ref{Network_node}(b)). $ISI$ are not constant and the number of small amplitude spikes between two large amplitude consecutive spikes is also varied in $SAO$s. The neighbors of this node have two silent (blue) and five oscillatory nodes (red). The number of spikes is slightly increased for another neuron originally in a quiescent state (Fig. \ref{Network_node}(c)) and the number of small amplitude spikes in $LAO$s is varied from $4$ to $5$. This neuron has 11 neighbors in which 7 nodes are self-oscillatory (blue) in the absence of coupling.

Next, we define the parameter $r_i$ to search for the presence of oscillatory nodes in the neighborhood of quiescent node ($i$) by
\begin{align}
r_i&=\frac{N_{oi}}{\sum_{j=1}^{N} A_{ij}} = \frac{N_{oi}}{S_i},\label{r_imp}
\end{align}
where $N_{oi}$ is the number of spiking oscillators connected with the $i$th quiescent node and $S_i$ the degree of the $i$th node. The neighbors of a third selected node are all oscillatory ($r=1$) and the node reveals lower $ISI$ as there is comparably fast switching from $SAO$s to $LAO$s (see Fig. \ref{Network_node}(d)). Therefore, the ratio of adjacent spiking nodes (blue) with respect to neighbors, $S_i$, determines the effect of the average $ISI$, $\langle ISI\rangle$, on the $i$th quiescent node (red). To understand the effect of the average $r$ on $\langle ISI\rangle$, we have considered three couplings: $K=0.3$, 0.4 and 0.6, shown in Fig. \ref{Network_node}(e) with upper red line (filled circle), middle red line (filled diamond) and lower red line (star), respectively. For the weaker couplings $K=0.3$ and $K=0.4$, and for small $r$, $\langle ISI\rangle$ exhibits significantly higher values (25 time units with high fluctuations). For higher values of $r\approx1$, $\langle ISI\rangle$ is decreased by 10 time units. The results confirm that, a red node with smaller $r$ (where the presence of red (quiescent) neighbors is significantly larger, have strong impact on the red node) reduces the number of spikes compared to the case where $r\approx 1$. For even higher couplings ($K=0.6$, red line with star marker), $\langle ISI\rangle$ decreases to around 5 and the impact of $r$ on$\langle ISI\rangle$ is not prominent at even higher couplings (not shown herein). We note that, as we have seen in Fig. \ref{Network_node}(b)-(d), smaller changes in $r$ ($r=\frac{2}{7}\approx0.28$, $r=\frac{7}{11}\approx0.63$ and $r=1$ for (b), (c) and (d), respectively) result in small amplitude spikes in $SAO$s between two large amplitude  spikes ($LAO$s). $\langle ISI\rangle$ and spikes in $SAO$s of quiescent nodes are determined by two key factors: the number of neighboring spiking neurons and the coupling strength. Therefore, we conclude that $\langle ISI\rangle$ decreases if the number of oscillatory nodes in the neighbour increases.

\vspace{1cm}
\subsubsection{MMOs of quiescent nodes: the role of electrical coupling}\label{imp_electricalcoup}

Next, we choose randomly a quiescent node (red) and check the effect of electrical coupling strength on MMOs connected to that node. At the lower coupling $K=0.3$, the node exhibits three small amplitude spikes ($SAO$s) between two large amplitude  spikes (Fig. \ref{MMO_weakcoupling}(b)). To quantify the spike distribution, we define 
\begin{align*}
f_{SAO}&=\frac{S_{SAO}}{S_{all}},\\
f_{LAO}&=\frac{S_{LAO}}{S_{all}},
\end{align*} 
where $S_{SAO}$, $S_{LAO}$ are the numbers of small and large amplitude  spikes, respectively, and $S_{all}$ the count of all spike amplitudes in the same interval. In Fig. \ref{MMO_weakcoupling}(b), three small amplitude spikes appear consecutively and are shown by star, triangle and hexagon markers, respectively. They are distributed with almost similar amplitudes (see left part of Fig. \ref{MMO_weakcoupling}(a) shown in light blue). As the membrane voltage is periodic, $f_{LAO}$ shares almost equal probability with $f_{SAO}$. We note that, we have used $f$ in Fig. \ref{MMO_weakcoupling}(b) instead of $f_{SAO}$ or $f_{LAO}$ to accumulate the information of the entire spiking frequency set. If we increase the coupling to $K=0.4$, we see that three small amplitude spikes converge to a single one (Fig. \ref{MMO_weakcoupling}(c), diamond marker), the oscillatory neighbors influence the oscillation of the quiescent node and they are equiprobable (the light and deep blue bars in Fig. \ref{MMO_weakcoupling}(a) are almost of the same amplitudes). At $K=0.6$, the small amplitude  spikes appear recurrently (circle marker in Fig. \ref{MMO_weakcoupling}(d)) after two large amplitude  spikes and give rise to MMBOs. Interestingly, simple MMOs change into more complex dynamics, i.e., MMBOs. Therefore, $f_{LAO}$ (deep blue bar) is higher than $f_{SAO}$ for small amplitude  spikes (light blue bar). When the coupling is set to $1$, the MMOs are completely lost (no light blue bar appears in the right-hand side of Fig. \ref{MMO_weakcoupling}(a), see also the spiking behaviour in Fig. \ref{MMO_weakcoupling}(e)). The quiescent neighbours at weak coupling contribute strongly to the generation of mixed-mode oscillations. When we increase the coupling, more information is shared among nearest neighbour nodes and long distant neighbors. The dynamics in the network, including that of quiescent nodes, is characterised by large amplitude  spikes. We note that, the nodes in the random network are dominated by self-oscillatory neurons ($70\%$) and for higher coupling, they control the spiking behavior in the entire network, therefore quiescent nodes cannot reflect MMOs for higher couplings.
\\\\
\subsubsection{Average $ISI$ vs coupling strength $K$ in neural subpopulations}

Here, we scan the average $ISI$, $<ISI>$, interval of the entire subpopulation varying the coupling strength $K$. The $\langle ISI\rangle$ of oscillatory (blue) nodes in the network is slightly increased (see Fig. \ref{ISI}(a) with filled blue circles) for weaker couplings and saturates around 5.6 time units when it is increased (for $K>1.2$). On the other hand, the $\langle ISI\rangle$ of red quiescent nodes is decreased when the coupling is increased. For small couplings, $\langle ISI\rangle$ shows strong fluctuations (shown by black lines with error bars in the backdrop of red filled circles, Fig. \ref{ISI}(b)) due to the desynchronized $\langle ISI\rangle$ in MMOs of the quiescent nodes. The red and blue lines in Fig. \ref{ISI}(a),(b) are plotted from the two coupled reduced models derived from the collective behavior of the connected network described in the next section. For small couplings, we see that the $\langle ISI\rangle$ of each quiescent node are dissimilar (see Fig. \ref{Network_node}), i.e., the firing rate varies from one node to another. We scan the entire average $ISI$ interval of the quiescent subpopulation for a range of coupling strengths to understand the fluctuations in $ISI$. To quantify these fluctuations, we calculate the coefficient of variation, $CV$, of $ISI$ of the quiescent subpopulation calculated from the numerical data (Fig. \ref{ISI}(c), red line with dots). $CV$ becomes zero after a certain coupling strength, as there is no variation in spike sequences and $SAO$s are completely vanished. The brown line in Fig. \ref{ISI}(c) reflects the frequency of peaks in the $SAO$s, which is zero for higher couplings, where $CV$ is also zero, thus revealing a close relation between $CV^2$ and $f_{SAO}$. In the Supplementary material, we present an analytical approach that relates the two quantities and offer a plausible explanation for the discrepancy observed for small coupling strengths.
 
{

\section{REDUCED MODEL DESCRIPTION}\label{reduce network}

 It is clear from Fig. \ref{TS_Spatio1} that the neurons within the subpopulations are synchronized for higher couplings, {\color{red} and therein} cluster synchronization appears within a subpopulation. This motivates us to pursue further an approach to construct a reduced model of two coupled systems which is able to encode the information in the large network. Since we have considered a random network in which the node-degrees follow the Poisson distribution, we can approximate the degree of each node/neuron by the average degree of the considered network \citep{Hens2015,Sasai2015}. Therefore, we can assume that
$S_j=\langle S\rangle$ for $j=1,\ldots,N$. The number of spiking oscillators in the neighborhood of each oscillator is expected to be $(1-\frac{p}{N})S=\frac{q}{N}S$
and that of quiescent oscillators, $\frac {p}{N}S$, where $p$ is the number of quiescent oscillators in the network. We set $v_j = V_Q$ for $j=1,\ldots,p$ and $v_l=V_S$ for $l=p+1,\ldots,N$. Over a certain coupling strength, within different clusters, the quiescent and spiking oscillators are synchronized separately. Therefore, by representing the two clustered subpopulations by two nodes, we obtain the following reduced system of coupled equations
\begin{align}
{\dot{V}}_{S}&=0.04{V}_{S}^2 + 5{V}_{S} + 140 - {U}_{S} + I_{S} + K p ({V}_{Q} - {V}_{S}),\label{reduce1}\\
{\dot{U}}_{S}&=a(b {V}_{S} - {U}_{S}),\label{reduce2}\\
{\dot{V}}_{Q}&=0.04{V}_{Q}^2 + 5{V}_{Q} + 140 - {U}_{Q} + I_{Q} + K q ({V}_{S} - {V}_{Q}),\label{reduce3}\\
{\dot{U}}_{Q}&=a(b {V}_{Q} - {U}_{Q}),\label{reduce4}
\end{align}
with the constraint equation that if
${V}_{Q} \geq 30$, 
then ${V}_{Q}\gets c$ and ${U}_{Q}\gets {U}_{Q}+d$. These conditions are also valid for spiking nodes, i.e., for
Eqs. \eqref{reduce1} and \eqref{reduce2} for spike oscillators with $I_{S}=10$ and for Eqs. \eqref{reduce3} and \eqref{reduce4} for quiescent oscillators with $I_{Q}=3$. We note that, for homogeneous networks, there will be no effect of the assortativity (degree-degree correlation) on MMOs or on collective firing states as the number of quiescent oscillators in the neighborhood of each oscillator will not be affected. The $\langle ISI\rangle$ plotted for $V_S$ and $V_Q$ as a function of $K$ is shown in Fig. \ref{ISI}(a),(b) with red and blue dots, respectively. The results almost match with the result for the random network (filled blue and red circles). A phase diagram of the coupled reduced model with respect to $\frac{p}{N}$ and $K$ is shown in Fig. \ref{phasespace}(a). The diagram is drawn by monitoring $V_Q$. The MMOs and spike regions are identified with the help of $f$ and quiescent (death) states by noting the variation of the peak values of $V_Q$. The dark-red regime is the steady state island, where all neurons in the random network remain in quiescent states. The regime of MMOs appears for weak couplings (for all $p$) shown in orange. The uncoupled quiescent nodes are desynchronized in this regime. All nodes collectively (and individually) fire at higher couplings for $p<0.9$ (pink region). The boundaries of each region are consistent with the results from the random network. To confirm further the onset of steady states, we have performed a bifurcation analysis to check the boundaries while we have changed $\frac{p}{N}$ from 0.8 to 1 for coupling strengths $K=2$ and $K=3$, respectively (see Fig. \ref{phasespace}(b),(c)). The stable fixed point, $V_Q$, is shown with thick green line in both cases. This fixed point (node) collides with a saddle point and vanishes at $\frac{p}{N}\approx 0.87$. The system shows spiking oscillations below $\frac{p}{N}\approx 0.87$	in both cases. Finally, for $\frac{p}{N}=0.95$, the system changes its dynamics from MMOs to a steady state at $K\approx 0.77$, as evidenced in Fig. \ref{phasespace}(d).

\section{EMERGENCE OF MMOs IN A SMALL-WORLD NETWORK}\label{MMO in SMW}
Following up the previous studies on a random network of neural computation, we construct here a small-world network of $N=500$ nodes. A closed non-local ring is constructed with $8$ adjacent neighbours. A rewire strategy \citep{Watts1998Smallworld} is implemented with a probability $0.2$ to construct the final network (see Fig. \ref{msallworld_reffect}(a)). To understand the impact of oscillatory neighbors (i.e., blue nodes) (see Eq. \eqref{r_imp}) on quiescent nodes (red), we have identified four quiescent nodes (red) with different 
$r$. 
The network comprises $40\%$ quiescent nodes. Nodes with higher percentage of oscillatory neighbors show spiking and irregular MMOs that appear between two successive spikes (Fig. \ref{msallworld_reffect}(b),(e), where $r=0.75$ and 1, respectively). However, the red nodes with less percentage of oscillatory neighbors are unable to fire ($r\approx0.4$, Fig. \ref{msallworld_reffect}(c)) or irregular spikes appear with higher $\langle ISI\rangle$ value ($r=0.5$, Fig. \ref{msallworld_reffect}(d)). The coupling strength is fixed at $K=0.3$. Figure \ref{msallworld_reffect}(e) shows the impact of $r$ on $\langle ISI\rangle$, which is seen to continuously decrease for nodes with large
percentage of oscillatory neighbors ($r\gg 0.1$). The average $\langle ISI\rangle$ saturates below 30 (red curve with black filled, circles) for $K=0.3$. For this coupling strength, diverse MMOs  can be seen in Fig. \ref{msallworld_reffect}(b)-(e). For the  higher coupling strength $K=0.4$, $\langle ISI\rangle$
 converges to 10 (red curve with black filled, diamonds). $r$ contributes less to $\langle ISI\rangle$ with the value fluctuating around 10 for $K=0.6$ (red curve with black filled, stars).

\section{Conclusions}\label{Conclusions}

In this paper, we sought to study MMOs in a random and a small-world network of diverse excitable Izhikevich neurons for different coupling strengths by introducing the generation of complex oscillations. We have observed  MMBOs, which are periodic in nature and are relevant to the  GnRH  model neuron  as the dynamical behavior of these neurons in a small-size network can be useful in the studies for epilepsy \citep{desroches2013mixed}. We have confirmed that a certain mixed population of quiescent and oscillatory nodes can give rise to several types of MMOs and MMBOs in the two types of networks.
 MMOs have potential applications in biophysical and other systems. In complex systems, various mechanisms exist during different oscillatory phases that generate spike patterns between fast and slow amplitude motion together with spikes and subthreshold oscillations, termed MMOs. It was observed that pyramidal neurons are capable of exhibiting two types of MMOs and their characterization was analyzed under antiepileptic drug conditions \citep{babak2017mixed}. Small amplitude oscillations ($<$10mV) give rise to intrinsic neuronal phenomena that exist during the synaptic transmission block \citep{alonso1989subthreshold,zemankovics2010differences}. Actually, it has been observed in many types of neurons such as in neurons in the thalamus, hippocampal CA1 neurons, neocortex neurons, spinal motor neurons, etc. \citep{puil1994resonant,gutfreund1995subthreshold,narayanan2007long,iglesias2011mixed}. It was suggested that MMOs can be responsible for the transition from high firing rates to quiescent states by reducing neuronal gain \citep{iglesias2011mixed,golomb2014mechanism}. Many studies showed the impacts of small amplitude oscillations/subthreshold oscillations (STOs) on diverse neuronal responses such as spike clustering \citep{puil1994resonant,gutfreund1995subthreshold,narayanan2007long}, synaptic plasticity \citep{narayanan2007long,bazzigaluppi2012olivary}, rhythmic activities, synchronization \citep{engel2008subthreshold,acker2003synchronization}, etc. 

Here,  random networks with various injected electrical current stimuli go through different transition phases of oscillations for various coupling strengths and emerging STOs with spikes, i.e., MMOs. First, the depolarization in membrane voltages show small amplitude oscillations around steady state potentials, and with further depolarization, gives rise to spikes, e.g. to MMOs \citep{jalics2010mixed}. STOs play an important role in the emergence of MMOs and in controlling spike clustering \citep{torben2012generation,latorre2016interplay}.

Furthermore, MMOs play an important role in neuronal functional mechanisms, namely, the STOs affect the sensitivity of neurons for injected input stimuli, the amplification of synaptic inputs and network synchronization to specific firing frequencies \citep{babak2017mixed}. The mechanism of MMOs produced in complex dynamical systems remains a challenging task. In the excitable pituitary cell model, pseudo-plateau bursting is canard-induced MMOs \citep{vo2010mixed}. It correlates electrophysiological behaviour of $SAO$s on clustering spikes, and shows the influences of ionic currents to the firing rate and spike patterns in the network.
Finally, experimental and numerical studies show that MMOs occur in oscillatory rhythms in brain functioning from a single neuron to global neural networks \citep{erchova2008rhythms}. In this study, we investigated both types of oscillations, MMOs and MMBOs. The results may be useful to Neuroscientists and those working on the mathematical modelling and dynamical behaviour of cortical neurons based in random neural networks.We plan in a future publication to explore the impact of excitatory and inhibitory connections in Izhikevich neurons and how they give rise to the emergence of MMOs \citep{noback2005human,deco2014local,pastore2018identification}.

\section*{Author Contributions}
CH and AM designed the research study and developed the results. SG, AM and CH designed the figures and SG performed the analytical and numerical simulations. CH, CGA and AM wrote the manuscript with support from SKD and PJ. SKD, AM, PJ and CGA provided support by constructive suggestions and feedback.

\section*{Funding}
 CH is supported by the INSPIRE-Faculty grant (code: IFA17-PH193).
\section*{Acknowledgments}
{\bf Conflict of Interest Statement:} The authors declare no competing interests.
\section*{Supplemental Data}



\subsection{Relation between ${CV}$ and $f_{SAO}$}


To understand the relationship between the coefficient of variation, $CV$ and $f_{SAO}$, we consider that the spikes in $LAO$s appear with probability $f_{LAO}$ and peaks in $SAO$s with probability $f_{SAO}=1-f_{LAO}$. 
Furthermore, we assume that we have a sequence of spike-time intervals as $\{T_{LAO},\dots,T_{SAO},\dots,T_{LAO}\}$. Based on the Bernoulli process \citep{golomb2014mechanism}, if $T_{LAO}$ appears with probability $f_{LAO}$ in the entire sequence, then $kT_{LAO}$ (where $k$ is an integer with $k\ge2$) will appear with probability $(1-f_{LAO})^{k-1}f_{LAO}$. Therefore,
\begin{align*}
\langle ISI\rangle_{n}&=\sum_{k=1}^{n} k T_{LAO}(1-f_{LAO})^{k-1} f_{LAO}\\
&=f_{LAO} \sum_{k=1}^{n} k T_{LAO}(f_{SAO})^{k-1}\\
&=f_{LAO} T_{LAO} \frac{d}{d(f_{SAO})} \sum_{k=1}^{n} \bigg( (f_{SAO})^{k}\bigg).
\end{align*}
Setting $f_{SAO}=x \in [0,1)$, we have that
\begin{align*}
\sum_{k=0}^{n} x^{k}&=\frac{1-x^{n+1}}{1-x}\\
&=\frac{1-x^{n+1}}{1-x}-1\\
&=\frac{x(1-x^{n})}{1-x}.
\end{align*}
Thus, $$\sum_{k=1}^{n} f_{SAO}^{k}=\frac{f_{SAO}(1-f_{SAO}^{n})}{1-f_{SAO}}.$$

Next, we compute $\langle ISI\rangle_{n}$
\begin{align*}
\langle ISI\rangle_{n}&=f_{LAO} T_{LAO} \frac{d}{d(f_{SAO})}\bigg(\frac{f_{SAO}(1-f_{SAO}^{n})}{1-f_{SAO}}\bigg)\\
&=f_{LAO} T_{LAO} \bigg(\frac{n(f_{SAO})^{n+1}-(n+1)(f_{SAO})^n +1}{(1-f_{SAO})^2}\bigg),
\end{align*}
and, in the limit of $n\to \infty$, i.e., $\lim_{n\to\infty}$, we have
\begin{align}\label{average_ISI_equation}
\langle ISI\rangle&=f_{LAO} T_{LAO} \frac{1}{(1-f_{SAO})^2}=\frac{T_{LAO}}{f_{LAO}},
\end{align}
where $f_{LAO}=1-f_{SAO}$.

Then,
\begin{align}\label{ISIsqaure}
\langle ISI^2 \rangle_{n}&=\sum_{k=1}^{n} k^2 T_{LAO}^2(1-f_{LAO})^{k-1} f_{LAO}\nonumber\\
&=f_{LAO} \sum_{k=1}^{n} k^2 T_{LAO}^2(f_{SAO})^{k-1}\nonumber\\
&=f_{LAO}T_{LAO}^2 \sum_{k=1}^{n} k^2 (f_{SAO})^{k-1}\nonumber\\
&=f_{LAO} T_{LAO}^2 \bigg(-\frac{d}{d(f_{SAO})} \sum_{k=1}^{\infty} (f_{SAO})^{k} +\frac{d^2}{d(f_{SAO})^2} \sum_{k=1}^{\infty} (f_{SAO})^{k+1}\bigg).
\end{align}
Manipulating Eq. \eqref{ISIsqaure} further, in the limit of ${n\to \infty}$, we get that
\begin{align}
\lim_{n\to \infty}\langle ISI^2 \rangle_n&=\langle ISI^2 \rangle\nonumber\\
&=f_{LAO} T_{LAO}^2 \bigg(\frac{2}{f_{LAO}^3}- \frac{1}{f_{LAO}^2}\bigg)\label{average_ISI^2_equation}.
\end{align}
Combining Eqs. \eqref{average_ISI_equation} and \eqref{average_ISI^2_equation}, we find that
\begin{align*}
CV&=\frac{\bigl(\langle ISI^2 \rangle-\langle ISI\rangle^2\bigr)^\frac{1}{2}}{\langle ISI\rangle}\\
&=\frac{ \bigg(f_{LAO} T_{LAO}^2 \Bigl(\frac {2} {f_{LAO}^{3}}-\frac{1}{f_{LAO}^2}\Bigr) - f_{LAO}^2 T_{LAO}^2 \frac{1}{f_{LAO}^{4}}\bigg)^{1/2}} {f_{LAO} T_{LAO} \frac{1}{f_{LAO}^{2}}}\\
&=\frac{ \bigg(\Bigl(-f_{LAO}^{-1} + 2 f_{LAO}^{-2}\Bigr) - f_{LAO}^{-2} \bigg)^{1/2}} { f_{LAO}^{-1}}\\
&=(1-f_{LAO})^{1/2}=(f_{SAO})^{1/2},
\end{align*}
thus, $$CV=f_{SAO}^{1/2},$$ where $CV\ge0$ and $f_{SAO}$ range in the interval $[0,1)$.

To validate our theoretical analysis, we have plotted $CV$ vs $\sqrt{f_{SAO}}$ in Fig. \ref{fig:s1} here for a wide range of couplings $K$ in $[0.0,2]$ . One can see that they follow a linear relationship. In particular, for higher coupling, $K \in [1, 2]$ both $CV$ and $\sqrt{f_{SAO}}$ tend to zero (near the origin in Fig. \ref{fig:s1}, see also Fig.\ 4(c) in the paper). However, for weak coupling (i.e., for $K$ in $[0,  1]$), these quantities deviate from each other and reside away from the origin (these points are depicted in the right top corner in Fig. \ref{fig:s1}, see also Fig. 4(c) in the paper). This ensures the existence of MMOs. The discrepancy appears due to the small sample size used to compute them, as we have considered integer $k$ values in the calculations above. In the future, we plan to explore the possibility that $k$ assumes real values in $[0,\infty)$.

\section*{References}
\bibliography{test_SD_ca1}
\bibliographystyle{apsrev4-1}



\begin{figure}[h!]
\begin{center}
\includegraphics[width=17.5cm,height=14cm]{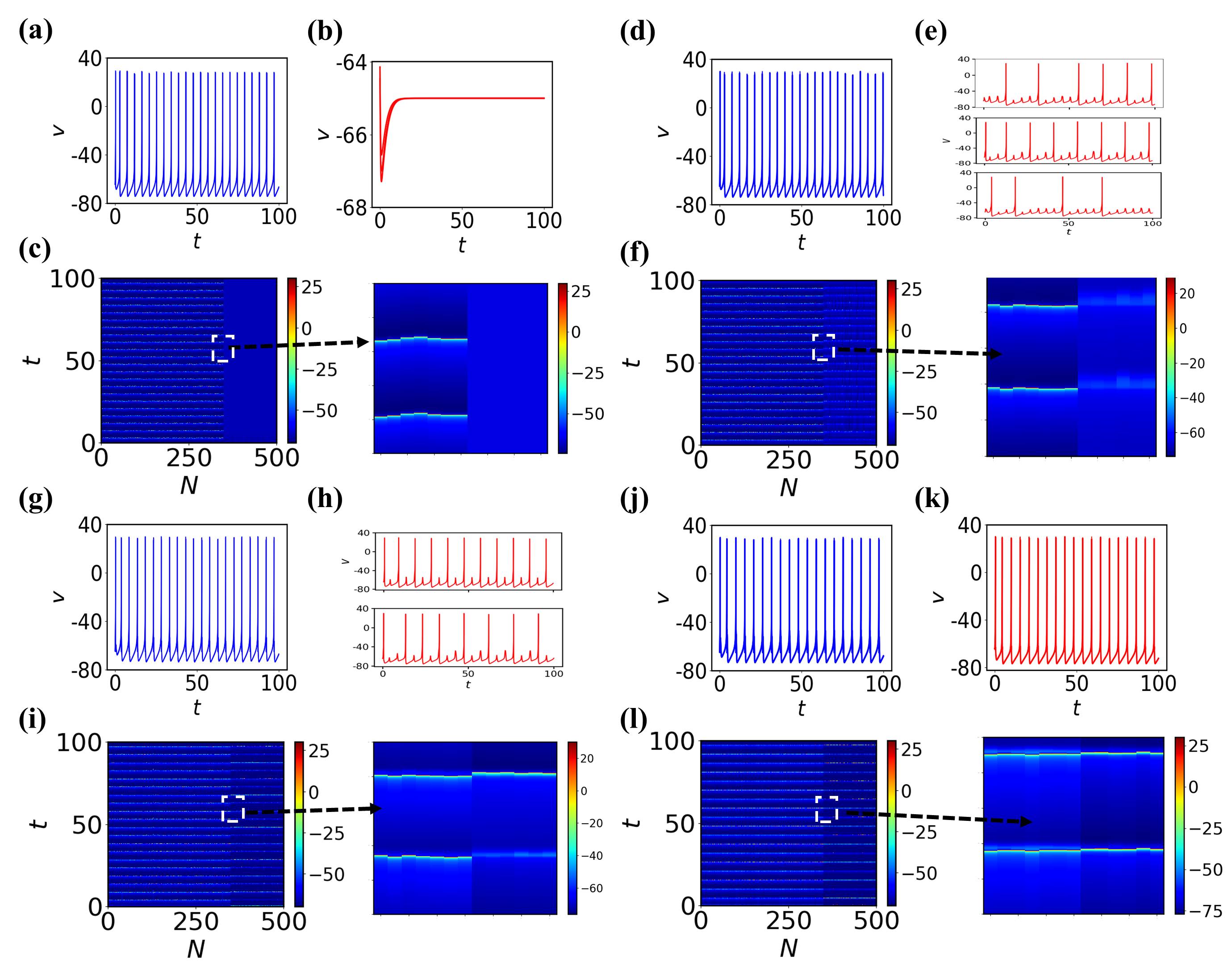}
\end{center}
\caption{\textbf{Membrane potential \boldmath{$v$} and spatiotemporal plots.} (a) One self-oscillatory spiking neuron in the absence of coupling ($K=0$) and a time-series of a quiescent node is shown in (b). (c) The spatiotemporal plot for all neurons in the random network. The first 350 nodes are self-oscillatory. Nodes from 351 to 500 are in steady states (see the 4 zoom-ins). (d),(e) The coupling is increased to $K=0.3$. There are several types of MMOs observed in the quiescent subpopulation. Three nodes from the quiescent subpopulation are marked and the {time series of each node over the course of time is shown in (e).} (f) Spatiotemporal plot of all neurons in the random network. The quiescent nodes are desynchronized with each other. (g),(h) The coupling is increased to $K=0.4$. $ISI$ of spiking nodes are increased and decreased for quiescent nodes. Desynchronized MMOs ({shown in (h), where two quiescent nodes have been randomly chosen}) are still visible in the quiescent population. (i) Spatiotemporal plot that shows the variation in spikes for all nodes in the random network. (j),(k) and (l) are for $K=1$. The entire population fires (without any MMOs appearing) with almost the same frequencies. Clearly two subpopulation are separately synchronized.}\label{TS_Spatio1}
\end{figure}


\begin{figure}[h!]
\begin{center}
\includegraphics[width=17.5cm]{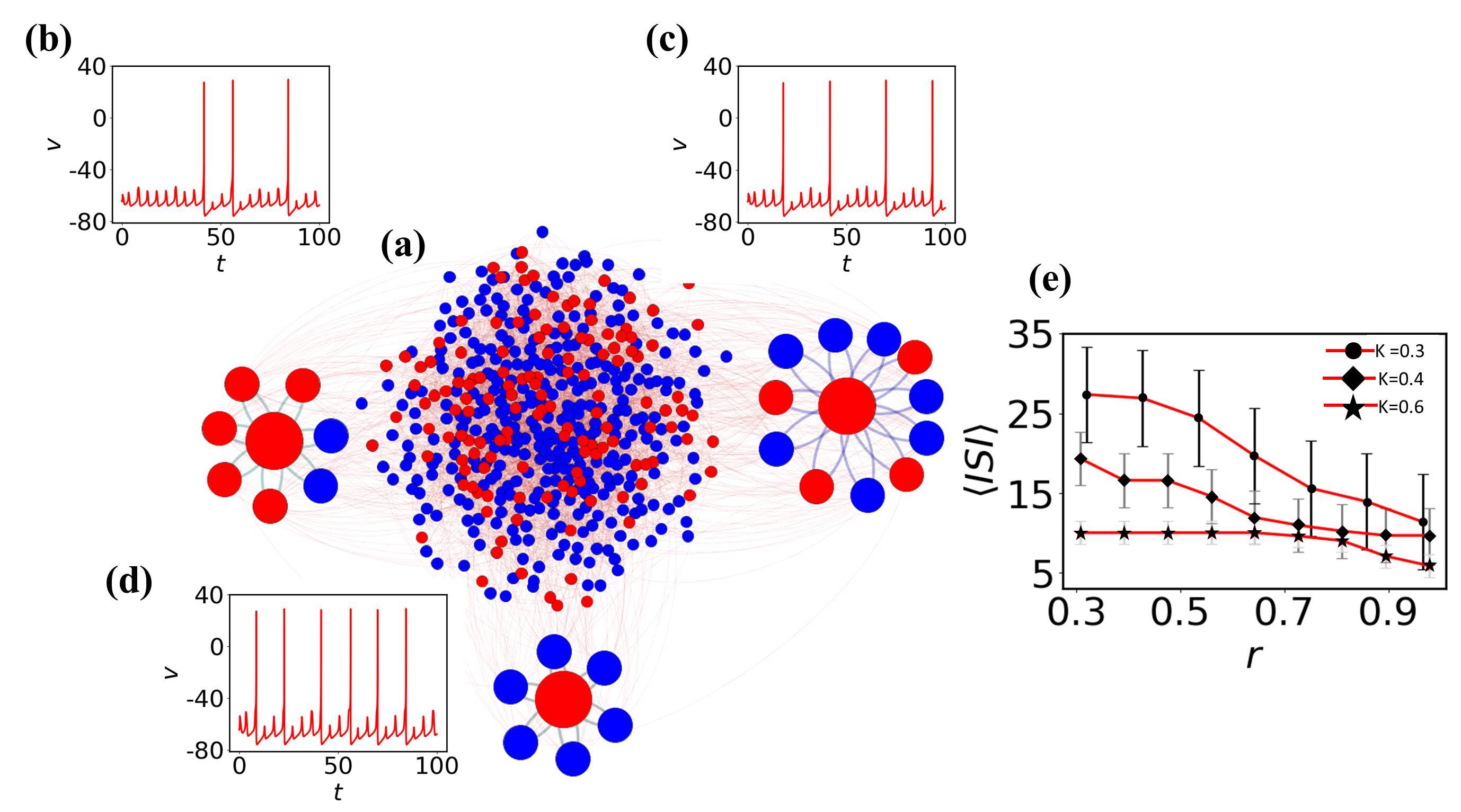}
\end{center}
\caption{{\bf The impact of neighbors of MMOs on quiescent nodes}. (a) The random network of 500 nodes \citep{Bastian2009}. Red nodes are in quiescent and blue in self-oscillatory states. (b) One red node is identified with degree 7. Five of them are spiking oscillators ($r\approx0.28$). Irregular MMOs are observed here. (c) The second red node with $r\approx0.63$. MMOs with considerably lower $ISI$ are shown. (d) All neighbors are self-oscillatory ($r=1$), MMOs with highly frequent spikes are observed. For (b)-(d), the coupling strength is fixed at $K=0.3$. (e) Impact of $r$ on $\langle ISI\rangle$. The $\langle ISI\rangle$ is continuously decreased if we check for higher values of $r$ and the average value saturates below 15 (red
	curve with black filled, circles, red curve with black filled, diamonds) for $K=0.3$ and $0.4$, respectively. For even higher coupling ($K=0.6$, red curve with black filled, stars), $r$ contributes less to $\langle ISI\rangle$ with the value fluctuating between 5 and 10.}\label{Network_node}
\end{figure}

\begin{figure}[h!]
\begin{center}
\includegraphics[width=17.3cm]{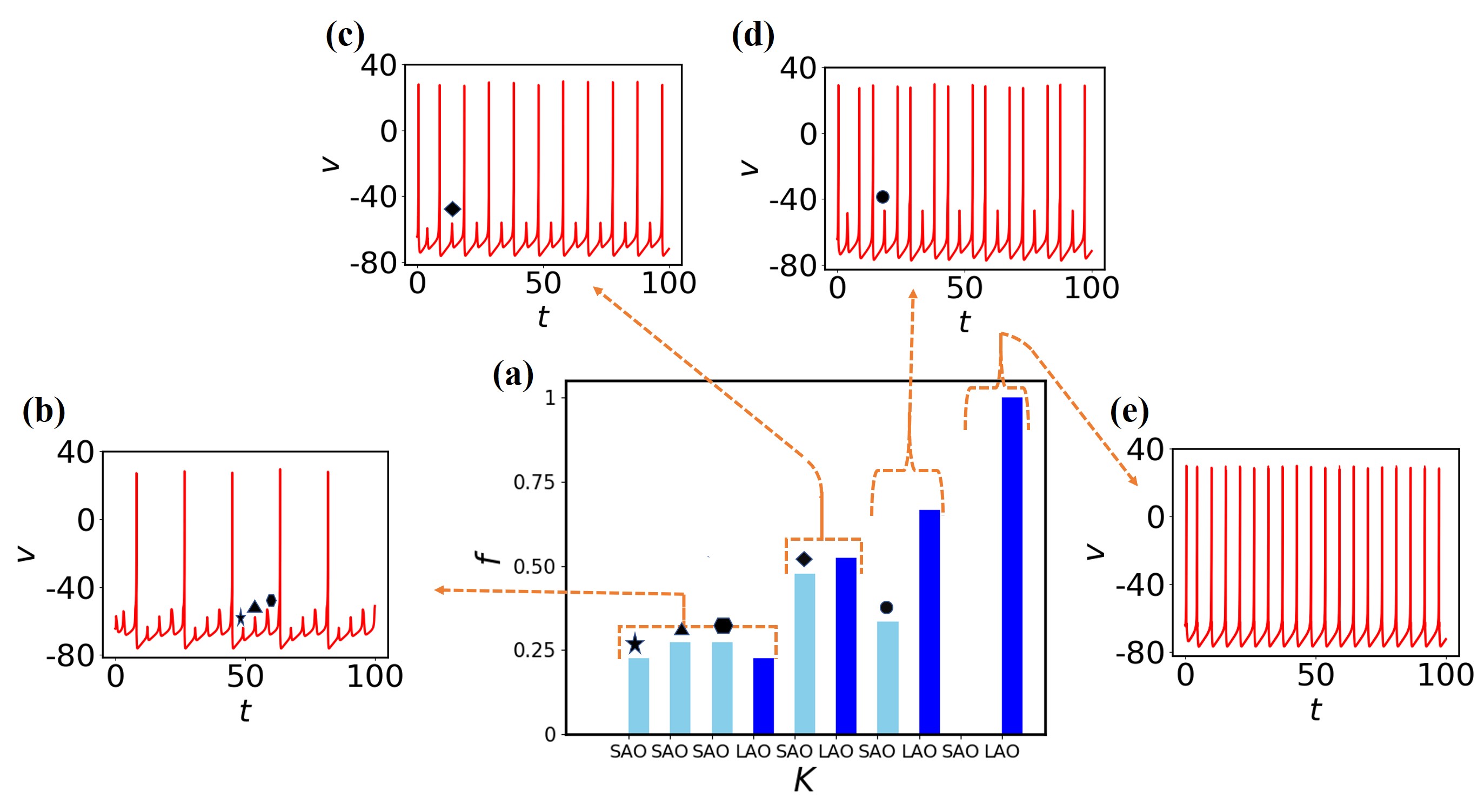}
\end{center}
\caption{{\bf Impact of coupling \boldmath{$K$} on MMOs of a quiescent (red) node.} (a) Probability distribution of spikes in $SAO$s (light blue) and $LAO$s (deep blue) for $K=0.3$, 0.4, 0.6 and 1 from left to right, respectively. (b) The time evolution for $K=0.3$. Three small amplitude oscillations (star, triangle and hexagon) appear between two consecutive large amplitude  spikes. (c) One small amplitude spike (diamond) appears between two large amplitude  spikes at $K=0.4$. (d) One small amplitude spike (black circle) appears after two spikes emerging together for $K=0.6$. Therefore, the probability of small amplitude spikes is decreased (third part of (a)) and results to the emergence of MMBOs. (e) Small spikes vanish at higher coupling ($K=1$), therefore MMOs are lost and tonic spikes are generated, instead.}\label{MMO_weakcoupling}
\end{figure}

\begin{figure}[h!]
\includegraphics[width=18cm]{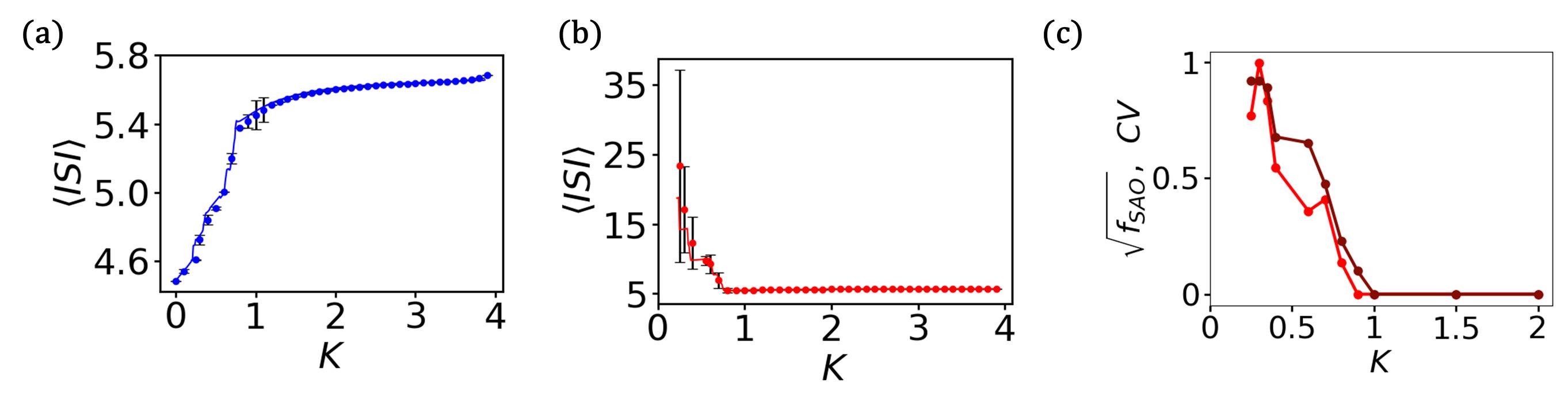}
\caption{\textbf{\boldmath{$\langle ISI\rangle$}, \boldmath{$\sqrt{f_{SAO}}$} and \boldmath{$CV$} as a function of coupling \boldmath{$K$}.} (a) $\langle ISI\rangle$ for all spiking oscillators (in total $350$). At small coupling, $\langle ISI\rangle$ is smaller, i.e., the spike frequencies are comparatively higher and it saturates around 5.6 for higher couplings. The fluctuations are negligible here, i.e., all spiking nodes have common frequencies for all couplings considered. (b) Quiescent nodes. For small couplings, the nodes exhibit diverse desynchronized MMOs (shown in black, with error bars). $\langle ISI\rangle$ saturates at higher couplings. (c) Relation between $CV$ (red line with marker) and $\sqrt{f_{SAO}}$ (brown line with marker) as a function of the coupling strength $K$.}
\label{ISI}
\end{figure}

\begin{figure}[h!]
\begin{center}
\includegraphics[width=17cm]{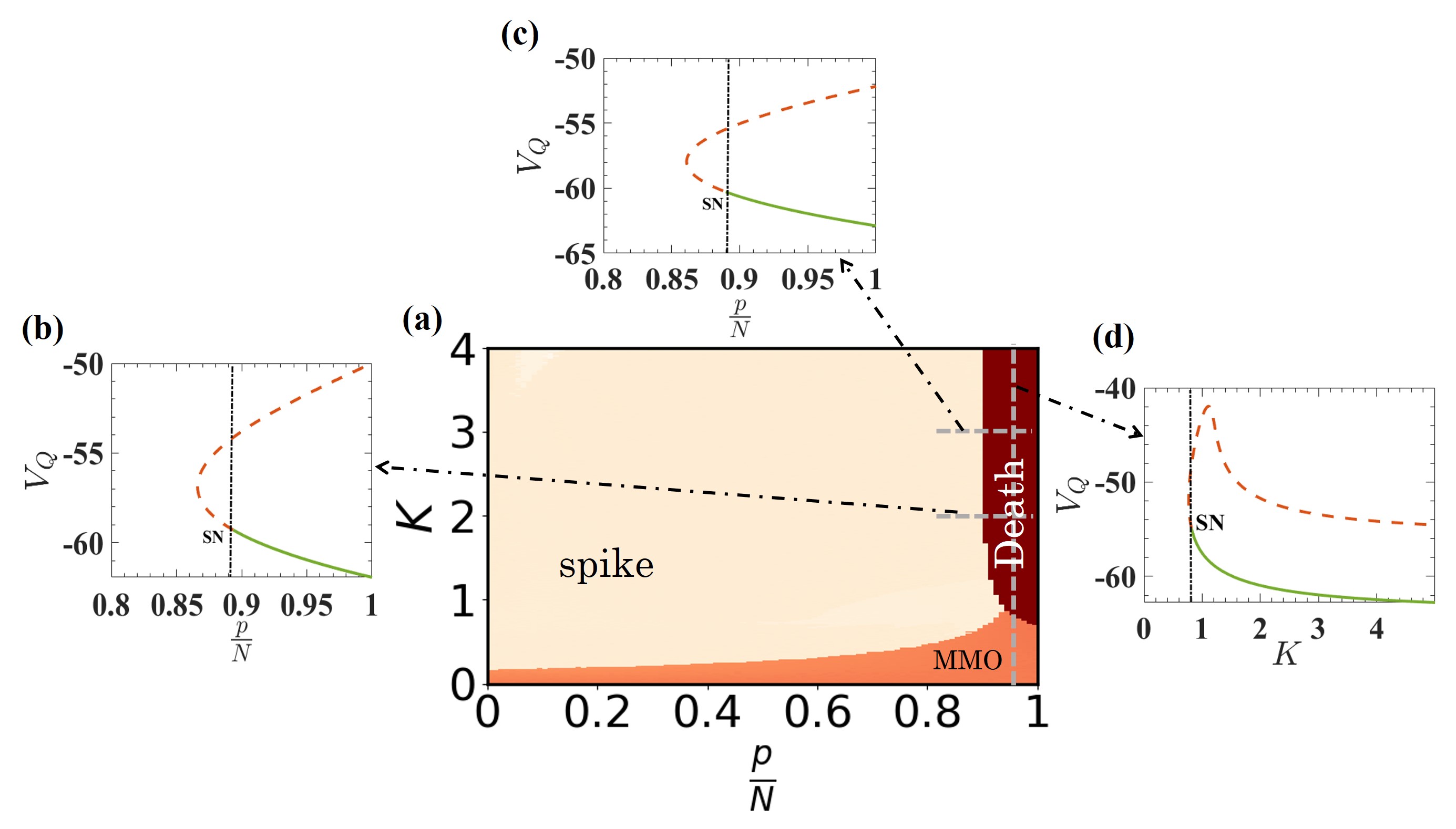}
\end{center}
\caption{\textbf{Phase-space diagram of the reduced quiescent node model as a function of \boldmath{$K$} and relative size of quiescent oscillators in the random network.} The emergence of MMOs, synchronized spiking oscillations and quiescent states are depicted in orange, pink and dark red, respectively. The boundaries of quiescent states with other regimes are demarcated by the bifurcation scenaria. (b),(c) Stable fixed points vanish through a saddle-node (SN) bifurcation at $\frac{p}{N}\approx 0.87$ for $K=2$ and 3, closely matched with the phase diagram. Note that for higher couplings, the boundary of quiescent states does not depend on $\frac{p}{N}$. (d) Bifurcation analysis as a function of $K$, for $\frac{p}{N}=0.95$ (dashed vertical line in (a)). The onset of quiescent states occurs at $K\approx 0.77$.}\label{phasespace}
\end{figure}

\begin{figure}[h!]
\begin{center}
\includegraphics[width=17cm]{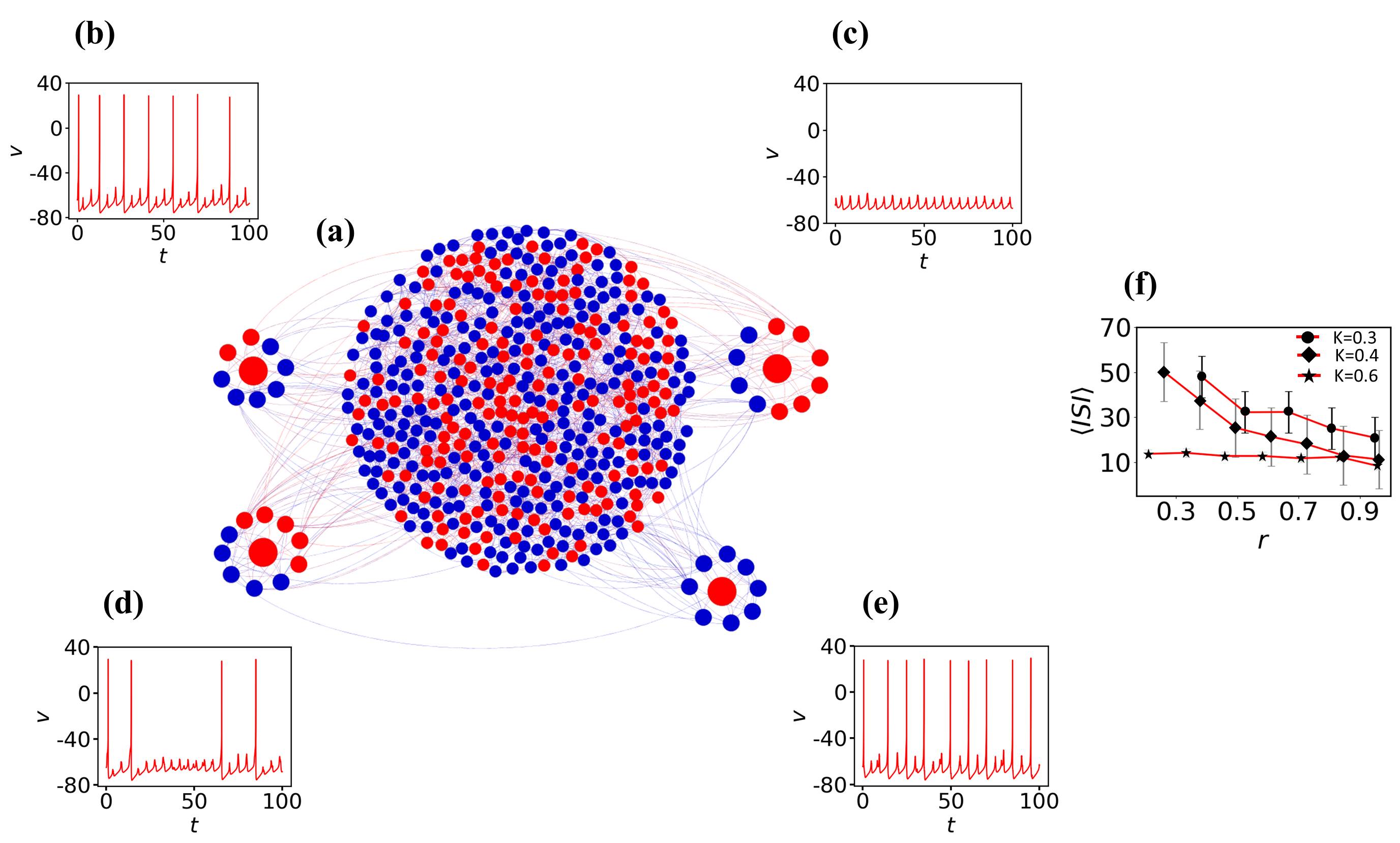}
\end{center}
\caption{\textbf{The impact of neighbors of MMOs on quiescent nodes}. (a) The small-world network of 500 nodes \citep{Watts1998Smallworld} with $p=0.2$ and $\langle S\rangle=8$. (b) One red node (quiescent) is identified with node-degree 8. Six of them are spiking oscillators ($r=0.75$). Irregular MMOs are observed here. (c) The second red node with $r\approx0.4$. The node shows sub-threshold oscillations only. (d)$50\%$ of the neighbor nodes are spiking oscillators and irregular spikes appear with high $\langle ISI\rangle$. (e) All neighbors are self-oscillatory ($r=1$) and MMOs with highly frequent spikes are observed. For (b)-(e), the coupling strength is fixed at $K=0.3$. (f) Impact of $r$ on $\langle ISI\rangle$. The $\langle ISI\rangle$ is continuously decreased if we increase $r$. The average value saturates below 30 (red curve with filled circles) for $K=0.3$ and converges to 10 (red curve with black filled,
	diamonds) for $K=0.4$. $r$ contributes less to $\langle ISI\rangle$ with the value fluctuating around 10 for $K=0.6$ (red curve with black filled, stars).}\label{msallworld_reffect}
\end{figure}

\begin{figure}[htbp]
	\begin{center}
		\includegraphics[scale=0.5]{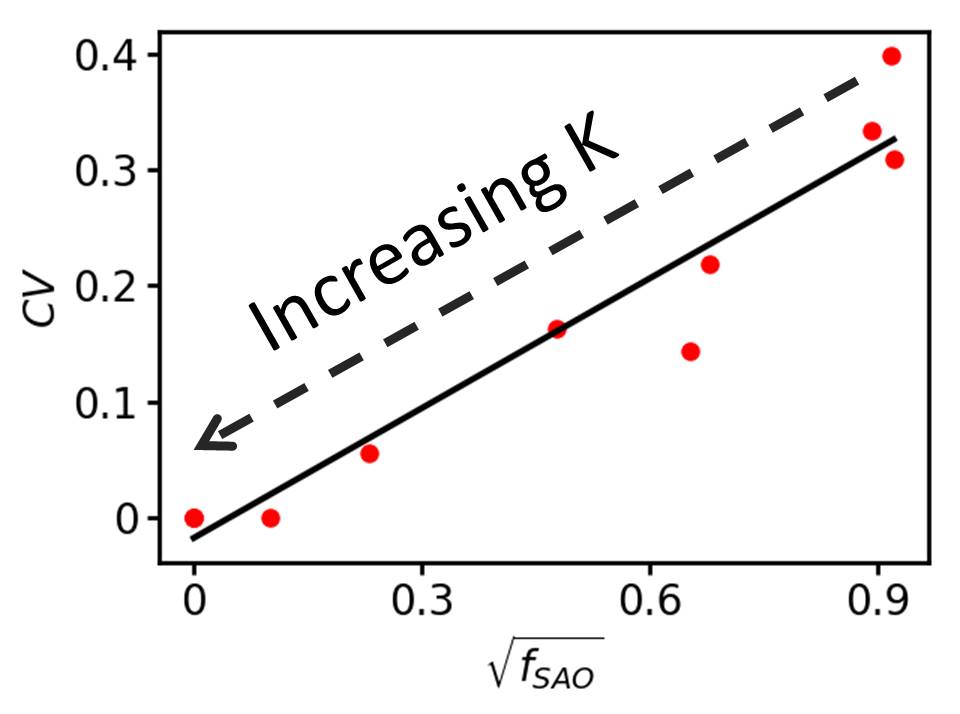}
	\end{center}
	\caption{\textbf{Linear relation between \boldmath{$CV$} and \boldmath{$\sqrt{f_{SAO}}$}.} The coupling strength $K$ is varied in $[0, 2]$ and the arrow shows the direction of increasing $K$ in $[0, 2]$. 
	}\label{fig:s1}
\end{figure}

\end{document}